\documentclass[twocolumn,aps,prb,reprint]{revtex4-1}
\usepackage[T1]{fontenc}
\usepackage[utf8]{inputenc}
\usepackage{xcolor}
\usepackage{graphicx}
\usepackage{amsmath}
\usepackage{xcolor}
\usepackage{ulem}
\usepackage[colorlinks=true, urlcolor=blue, citecolor=blue, linkcolor=blue]{hyperref}

\newcommand{\doiit}[1]{
	{\href{http://doi.org/#1}{http://doi.org/#1}}
}

\begin{document}

\def\Vhrulefill{\leavevmode\leaders\hrule height 0.7ex depth \dimexpr0.4pt-0.7ex\hfill\kern0pt}
\renewcommand{\vec}[1]{\mathbf{#1}}
\newcommand{\ii}{\mathrm{i}}

\title{Many-body localization phase in  a spin-driven chiral multiferroic chain}
\author{S. Stagraczy\'nski$^1$, L. Chotorlishvili$^1$, M. Sch\"uler$^{1,2}$, M. Mierzejewski$^3$,  J. Berakdar$^1$}
\affiliation{
	$^1$Institut f\"ur Physik, Martin-Luther-Universit\"at Halle-Wittenberg, 06099 Halle, Germany\\
	$^2$Department of Physics, University of Fribourg, 1700 Fribourg, Switzerland\\	
	$^3$Institute of Physics, University of Silesia, 40-007 Katowice, Poland
}
\date{\today}

\begin{abstract}
	Many-body localization~(MBL) is an emergent phase in correlated quantum systems with promising applications, particularly in quantum information. Here, we unveil the existence and analyse this phase in a chiral multiferroic model system. Conventionally, MBL occurrence is traced via level statistics by implementing a standard finite-size scaling procedure. Here\textcolor[rgb]{0.8,0.4,0.0}{,} we present an approach based on the full distribution of the ratio of adjacent energy spacings. We find a strong broadening of the histograms of counts of these level spacings directly at the transition point from MBL to the ergodic phase. The broadening signals reliably the transition point without relying on an averaging procedure. The fast convergence of the histograms even for relatively small systems allows monitoring the MBL dynamics with much less computational effort. Numerical results are presented for a chiral spin chain with a dynamical Dzyaloshinskii Moriya~\textcolor[rgb]{0.0,0.0,0.0}{(DM)} interaction, an established model to describe the spin excitations in a single phase spin-driven  multiferroic system. The multiferroic MBL phase is uncovered and it is shown how to steer it via electric fields.
\end{abstract}

\maketitle

\section{Introduction}
	Disorder may localize propagating waves. This phenomenon, first unraveled for  noninteracting electronic systems~\cite{Anderson} applies also in a general setting involving even classical waves~\cite{wave1,wave2,wave3}. For the ground state of a correlated system disorder is also of a great importance and may inhibit conductance~\cite{eeanderson}. Between these two regimes interesting phenomena emerge as well: For a certain disorder strength, a correlated excited system may form a \textcolor[rgb]{0.0,0.0,0.0}{MBL phase}~\cite{Basko,Devakul,Luitz,Vasseur,Bardarson,Serbyn,topical} with distinct properties such as boundary-law entropy, localization protected order, and vanishing DC~(even non-linear) conductivity with MBL mobility edge between the localized and the thermalized~(we call it also ergodic) phase. There are indications of the quantum nature of MBL~\cite{huse}. Experiments on ultra-cold atoms in optical lattices~\cite{mblatoms1,mblatoms2,mblatoms3} and trapped ions~\cite{mblions} provided evidence of MBL.  Here, we propose studying the signature of MBL in  electromagnon excitations of a spin-driven multiferroic oxides \cite{chiral}. A possible MBL is expected to hinder the electromagnon transport which can be traced down for instance as in Ref.~[\onlinecite{bo}].
	One currently discussed question is how the spectrum of a system with MBL phase is distributed. For integrable systems this issue has long been in the focus of research with many interesting findings (cf.~for instance~\cite{Bertini,Rigol,Ilievski,Pozsgay,Mierzejewski,ETH1,ETH2,ETH3}). Generically, integrable systems with a large number of degrees of freedom turn  chaotic for a weak perturbation. For interacting systems, the MBL phase is possibly mappable onto a model with fewer degrees of freedom \cite{huse1,huse2} and hence is resistent to perturbations until reaching a critical perturbation strength destroying the MBL phase. This scenario can be assessed by indicators that signal the transition from the MBL to the thermalized~(ergodic) phase. Conventionally the levels statistics behavior is used for this purpose \cite{Bauer}. In the ergodic phase the level spacing follows a Wigner distribution function, while in the MBL phase it obeys a Poisson distribution.
	A key quantity is the disorder average of the ratio
	\begin{eqnarray}
		\label{statistics}
		 && r_n = \min \left( \delta_n, \delta_{n-1} \right) / \max \left(\delta_n, \delta_{n-1}\right),\nonumber\\
		 && r = \frac{1}{N-2} \sum\limits^{N}_{n=3} r_n,
	\end{eqnarray}
	where $\delta_n = E_n - E_{n-1}$ is the distance between two neighbor energy levels {labelled by $n$} and $N$ is the number of eigenstates.
	In the ergodic phase for the Gaussian orthogonal ensemble (GOE) $r_\texttt{{GOE}} = 0.5307$ is found \cite{Haake}, while in the MBL phase $r_\texttt{{Poisson}} = 0.3863$ . The disorder strength has a strong influence on the system's spectral characteristics. Typically the transition to the MBL phase sets in at a certain critical strength of disorder, as deduced from conventional finite-size scaling analysis, assumed to apply also to the MBL case~\cite{Luitz,Bardarson}. The spectrum is obtained via exact diagonalization. In this work we present a new method based on the histograms of counts (hereafter, histograms) of the inter-level distances. The transition point is marked by a broadening of the histograms and enhanced fluctuations.
%
\textcolor[rgb]{0.0,0.0,0.0}{
We will be dealing with a system with a mixed symmetry having the Hamiltonian (GUE denotes Gaussian unitary ensemble)
	\begin{equation}
		\hat{H}_{total}=\hat{H}_{\mathrm{GOE}}+\lambda\hat{H}_{\mathrm{GUE}},
		\label{eq:H_sym}
	\end{equation}
	 Systems with mixed GOE/GUE symmetries are of high interest and widely studied in the recent literature \cite{Bera}. Depending on the value of the parameter $\lambda$ the spectrum of the Hamiltonian~(\ref{eq:H_sym}) displays different features: In the limit of small $\lambda$ the  level statistics obey GOE, for large $\lambda$  GUE prevails.  Interestingly, in the  corssover regime the system shows qualitatively different properties than $\hat{H}_{\mathrm{GOE}}$ and $\hat{H}_{\mathrm{GUE}}$.  Particularly for such cases, further  methods to explore the MBL phase in  systems with mixed complex symmetries are useful and needed.
	 We observed that in spite of the difference of the GOE and GUE statistics, the enhanced fluctuations at the MBL transition point bear a  universal physical character common for  both GOE and GUE symmetries. Thus, our method serves as a useful tool to explore  MBL in systems  with complex symmetry properties.
}

\textcolor[rgb]{0.0,0.0,0.0}{
	The manuscript is structured as follows: In section~\ref{s:Hs12Hm}, we introduce the helical spin-1/2 Heisenberg model, in the section~\ref{s:Istp}, we explore the integrals and the  transformation properties of the helical spin-1/2 Heisenberg model. Section~\ref{s:MBLms} is dedicated to the MBL phase in the helical spin-1/2 Heisenberg model, and  in section~\ref{s:Qfhc}, we introduce the histograms of counts as a  tool for sensing the  MBL phase and explore the enhanced fluctuations near  the transition point.
}

\section{Helical spin-1/2 Heisenberg model}\label{s:Hs12Hm}
	MBL phenomena were analyzed for disordered quantum spin chains (cf.~for instance ~\cite{Luitz,Vasseur,Bardarson,Serbyn,topical}). Our focus here is on a particular system, namely on low-energy excitations of oxide-based, spin-driven multiferroics that were realized experimentally in a chain form \cite{chiral,Park,Schrettle}~(such as LiCu$_2$O$_2$), and are charge insulators. Such systems exhibit at low temperatures a helical spin order coupled to an effective electric polarization. The ferroelectric polarization order is described by~[\onlinecite{chiral}], $ {\bf P} = g_{ME} \sum\limits^L_{i=1}\left\langle \mathbf{e}_x \times \left( \mathbf{\hat S}_i \times \mathbf{\hat S}_{i+1} \right)\right\rangle $. Here, $g_{ME}$ is the strength of the magnetoelectric coupling. The {$L$} effective {spin}-$1/2$ moments localized at sites $i$ are described by the operators $\mathbf{\hat S}_i$ and $\vec{e}_x$ is the spatial unit vector along the $x$ axis connecting $\mathbf {\hat S}_i$ and $\mathbf {\hat S}_{i+1}$. Thus, the low energy excitations are electromagnons. A possible MBL phase has so a multiferroic nature and is controllable by a magnetic $\bf B$ field (we assumed it applied along the $z$ axis) or an electric field $\mathbf{E}$ (applied below along the $y$ axis) that couples to ${\bf P}$. Such a scenario is well captured by the low-energy, effective spin--$1/2$ Hamiltonian with a dynamical \textcolor[rgb]{0.0,0.0,0.0}{DM interaction}~\cite{chiral,Azimi}
	\begin{align}
		\label{Hamiltonian}
	  	\hat{H} &= J_1 \sum^{L}_{i=1}\hat{\vec{S}}_i \cdot \hat{\vec{S}}_{i+1} + J_2 \sum^L_{i=1}\hat{\vec{S}}_i\cdot \hat{\vec{S}}_{i+2}\\
	   &\quad+\sum^L_{i=1} B^z_i \hat{S}_i^z + D \sum^{L}_{i=1} \left(\hat{\vec{S}}_{i} \times \hat{\vec{S}}_{i+1} \right)_z,\; D=E_y g_{ME}.\nonumber
	\end{align}
	Note, $D$ combines the effect of the electric field and the magnetoelectric coupling. The nearest neighbor exchange interaction is ferromagnetic $J_1<0$ while the next nearest neighbor one is antiferromagnetic $J_2>0$ leading in general to a frustrated spin order. For $D=0$ {and uniform magnetic field $B^z_i = B^z$}, $\hat H$ is of the Majumdar-Ghosh type~\cite{Majumdar}. Breaking the SU$(2)$ symmetry, for instance by an intrinsic anisotropy, the expectation value of the vector chirality $\boldsymbol{\kappa}_i= \langle \hat{\vec{S}}_{i} \times \hat{\vec{S}}_{i+1} \rangle$ turns finite (see e.~g. [\onlinecite{Kecke2008},\onlinecite{Sirker}]), signaling the emergence of a spin ordering in the $xy$-projection. The dependence on $D$  implies a variation in $E_y$ for the material-specific $g_{ME}$. For $J_2=0$ the anisotropic Heisenberg chain is retrieved. We note that electromagnon excitations may soften the phonon modes due to different possible types of magnetoeleastic couplings such as exchange-striction~\cite{jia1,jia2}. This effect in relation to MBL is not addressed here. Generally, coupling MBL systems to an incoherent environment may restore ergodicity  washing out MBL signatures \cite{a,b,c,d}. Here, we qualitatively assess the role of phonon modes on MBL  by disordering $D$, as discussed below.
	Computationally we are able to deal with only  small chains as compared to experiment,  for instance on LiCu$_2$O$_2$ chain. On the other hand, our model is versatile and captures also the noncollinear spiral order evidenced experimentally in size-selected, individual bi-atomic Fe chains on the $(5\times 1)-$Ir(001) surface \cite{stm}.
\section{Integrals of the system and transformation properties}\label{s:Istp}
\textcolor[rgb]{0.0,0.0,0.0}{
	Our system $\hat H$ has certain integrals of motion. It is straightforward to show that   the total spin component $\hat{S}^{z}=\sum^L_{i=1}\hat{S}_i^{z}$ commutes with $\hat H$. Therefore, $\hat H$ is  block-diagonal.
}
	Each block is identified via the conserved total spin component $\hat{S}^{z}$. Of a special interest is the largest subspace of states $\left| \Psi_n \right\rangle$ { obeying $\hat{S}^{z} \left|\Psi_n\right\rangle=M \left|\Psi_n\right\rangle$ with $M=0$ for even $L$ or $M=1$ for odd $L$, respectively}. A uniform magnetic field $B^z_i = B^z$ shifts equally the eigenvalues in each subspace and has no prominent effect on the inter level distance $r_n$, while randomness incorporated in the magnetic field $B^z_i \in \left\langle -h, h \right\rangle$ can induce a qualitative change of the spectral properties from Wigner-Dyson to Poisson level spacing statistics. The strength of disorder is measured on a scale set by $J_{1}$. In what follows we work with dimensionless units such that \textcolor[rgb]{0.0,0.0,0.0}{$J_{1}=1$, $J_{2}\rightarrow J_{2}/J_{1}$, $D\rightarrow D/J_{1}$,
$B\rightarrow B/J_{1}$. For simplicity we omit the factor of 1/2 in front of spin operator $\hat{\vec{S}}\equiv \hat{\vec{\sigma}}$ implying an extra rescaling $J_{1,2}=J_{1,2}/4$, $D,B^{z}=D,B^{z}/2$}.
	Depending on the experimental realization different boundary conditions have to be taken into account. For open boundary conditions, a unitary local rotation of spins $\hat{S}_{k}^{+}\rightarrow\hat{S}_{k}^{+}e^{-\ii k\Theta}$, $\hat{S}_{k}^{-}\rightarrow \hat{S}_{k}^{-}e^{\ii k\Theta}$ around the $z$ axis by the angle $\Theta =-\arctan\big(D/J_{1}\big)$ converts the Hamiltonian~(\ref{Hamiltonian}) to
	\begin{eqnarray}
	 \label{Hamiltonian2}
	  \hat{H}_T = J_1 \sum^L_{i=1} \hat{S}_i^{z} \hat{S}_{i+1}^{z} + \frac{J^\prime_1}{2} \sum^L_{i=1}\big(\hat{S}_i^{+} \hat{S}_{i+1}^{-}+\hat{S}_i^{-} \hat{S}_{i+1}^{+}\big)\nonumber\\
	  J_2 \sum^L_{i=1}\hat{S}_i^{z} \hat{S}_{i+2}^{z} + \frac{J^\prime_2}{2} \sum^L_{i=1} \big(\hat{S}_i^{+} \hat{S}_{i+2}^{-}+\hat{S}_i^{-} \hat{S}_{i+2}^{+}\big)\nonumber\\
	   - \sum^L_{i=1} B^z_i \hat{S}_i^z  - D^\prime \sum^L_{i=1}\left(\hat{\vec{S}}_{i} \times \hat{\vec{S}}_{i+2} \right)_z.
	\end{eqnarray}
	Here $J^\prime_1 = \sqrt{J_1^{2}+D^{2}}$, $J^\prime_{2}=J_{2}\big(J_{1}^{2}-D^{2}\big)/\big(J_{1}^{2}+D^{2}\big)$, $D^\prime=DJ_{1}J_{2}/\big(J_{1}^{2}+D^{2}\big)$. As evident, for $J_{2}=0$ the \textcolor[rgb]{0.0,0.0,0.0}{DM} interaction term disappears and  the Hamiltonian is equivalent to the XXZ model; whereas for twisted periodic boundary conditions, a reminiscent of a \textcolor[rgb]{0.0,0.0,0.0}{DM interaction} remains~\cite{Bocquet}.

\section{Many-body localization with mixed symmetries}\label{s:MBLms}
\textcolor[rgb]{0.0,0.0,0.0}{
	For systems of mixed GOE/GUE symmetries~(\ref{eq:H_sym}) three asymptotic cases are of interest: a) $\hat{H}_{\mathrm{GOE}}$ term is the dominant term and $\hat{H}_{\mathrm{GUE}}$ is the small perturbation, b) Both terms $\hat{H}_{\mathrm{GOE}}$ and $\hat{H}_{\mathrm{GUE}}$ are of equal strength, c) $\hat{H}_{\mathrm{GUE}}$ term is the dominant term and $\hat{H}_{\mathrm{GOE}}$ is a small perturbation. The two cases a) and c) are relatively simple and well captured by a standard finite-size scaling procedure.  For case b)  the level statistics cannot be identified in terms of GOE and GUE. Systems of different size $L$ manifest non-equivalence and different features. The role of GUE enhances at larger $L$ (see below). In the numerically inaccessible limit $L\rightarrow\infty$ the system is characterized by GUE statistics. However, the collapse of different data for different numerically accessible finite $L$ to a single universal curve (as expected  by a finite-size scaling procedure) is generally not achievable. Here we focus on periodic boundary conditions.  Prior implementing a finite-size scaling procedure we present results for  systems of different lengths $L=9,10,..14$.}

\begin{figure}[t]
		\centering
		\includegraphics[width=1.0\columnwidth]{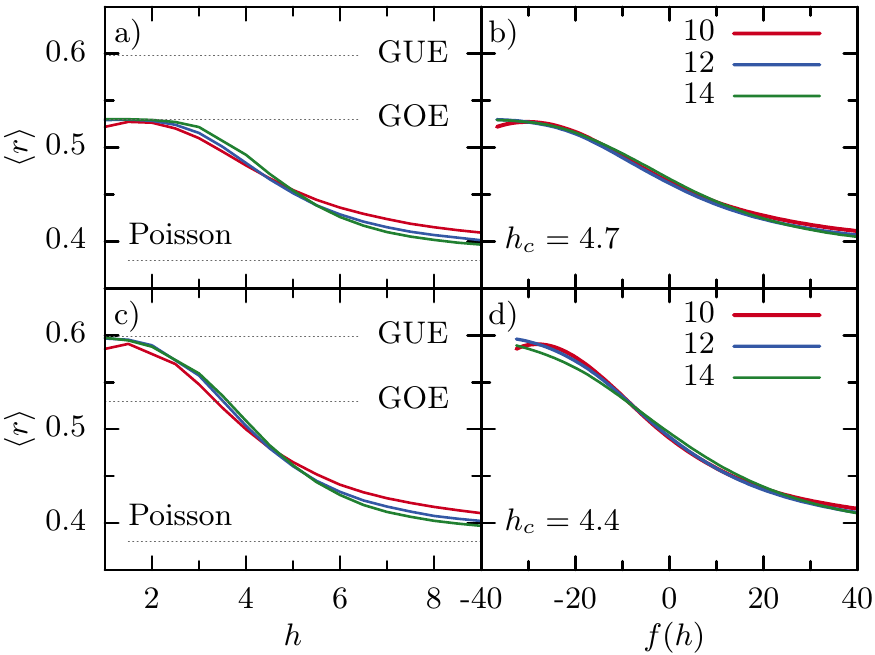}
		\caption{
	  		Level statistics for ferromagnetic nearest neighbor interaction $J_1$ = -1. For a)  $D = 0$. Increasing the randomness range $2h$ of the magnetic field $B^z \in  \left\langle -h, h \right\rangle$ the system spectrum undergoes a transition from  GOE to  Poisson distribution, indicating a tansition from ergodic to  MBL phase. Curves correspond to different system sizes and are obtained by an ensemble average for random disorder. b) Scaling collapse in the vicinity of the MBL transition is achieved by the  scaling function $f(h) = L^{1/\nu} (h - h_c)$, $\nu = 1$,  {$h_c = 4.7$}. c) With \textcolor[rgb]{0.0,0.0,0.0}{DM interaction} ($D = 0.2$). Periodic boundary conditions are implemented and the system is of GUE type. Increasing randomness the system undergoes a transition from GUE to the Poisson distribution. d) The rescaled phase transition curves from the ergodic to the MBL phase for $D = 0.2$, $h_c = 4.4$, $\nu = 1$, $f(h) = L^{1/\nu} (h - h_c)$.
	  	}
	  	\label{GOE_GUE}
	\end{figure}

\textcolor[rgb]{0.0,0.0,0.0}{
	Fig.~\ref{fig:DM0,00} shows the statistically averaged nearest neighbor inter-level distances for different strengths of disorder $h$ of the magnetic field $B^z \in \left\langle -h, h \right\rangle$. In particular, Fig.~\ref{fig:DM0,00} corresponds to the case a) when the symmetry of the system is precisely identified as  GOE and the DM term is zero. As inferred for zero field, the statistics obtained for systems of different lengths perfectly fits with a standard GOE $\left\langle r \right\rangle=0.53$. Increasing the strength of disorder $h$, the system performs a transition to the Poisson statistics and this transition grasps the concept of MBL phase. In the Fig.~\ref{fig:DM0,01} are shown the results for the case when DM term is small and GOE is the main symmetry of the system, while GUE is a perturbation. For this case we conclude that  the larger  the system size ($L=14$) the more important is the contribution of the DM term. Note that a finite-size scaling procedure relies on the collapse of different data for different $L$ to a single universal curve. Here we face a problem because for large $L$ the deviation from GOE is more prominent. For  stronger DM term the deviations from GOE  increase (see Fig.~\ref{fig:DM0,02}). For $D=0.05$ the system turns into mainly  GUE type and GOE is a perturbation see Fig.~\ref{fig:DM0,05}. However, this case still corresponds to case b), meaning a mixed complex symmetry and the finite-size scaling procedure fails. Only for unrealistically large $D=0.2$ we have the case c) (the $\hat{H}_{\mathrm{GUE}}$ term is dominant and $\hat{H}_{\mathrm{GOE}}$ is a small perturbation). For a strong DM interaction $D=0.2$ one observes a perfect switching from the GOE to the GUE level statistics and a finite-size scaling procedure provides reasonable results. (cf. Fig.~\ref{GOE_GUE}a and Fig.~\ref{GOE_GUE}c.) Thus, finite-size scaling is reliable when the DM interaction term is either zero or large enough and the symmetry of the system is mainly GOE or GUE, respectively.
}

	Each curve corresponds to a system with a different size and is obtained by means of an extensive averaging procedure over the ensemble of realizations of random disorder (up to 10.000 realizations per single point of the curve).
	\begin{figure}[ht]
		\includegraphics[width=0.9\columnwidth]{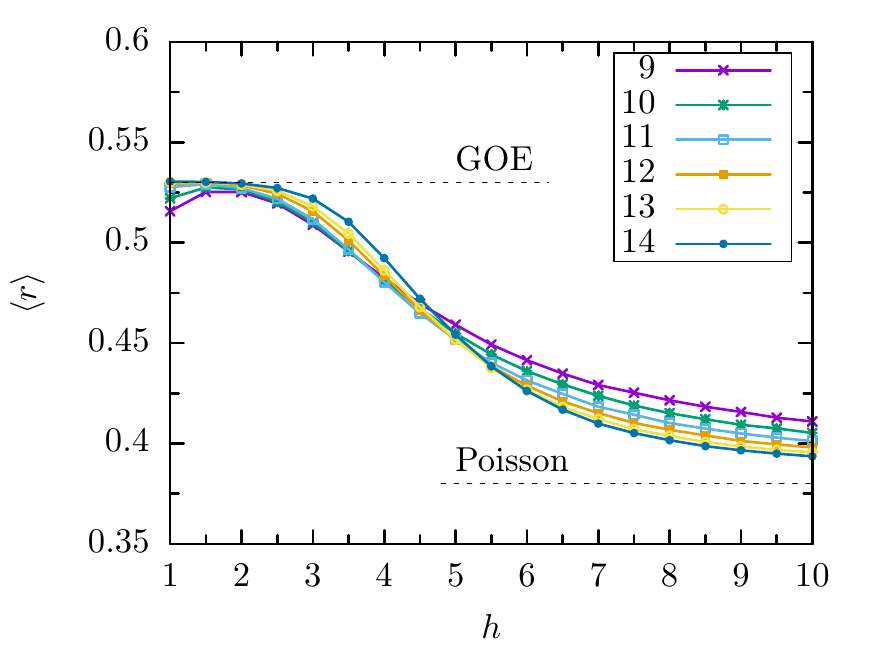}
		\caption{
		 \textcolor[rgb]{0.0,0.0,0.0}{
			Level statistics for a ferromagnetic nearest neighbor interaction $J_1 = -1$, as a function of the strength of disorder $h$ of the magnetic field $B^z \in  \left\langle -h, h \right\rangle$. The DM interaction term is zero and the system exhibits a  precise GOE symmetry.
		 }
		}
		\label{fig:DM0,00}
	\end{figure}
	\begin{figure}[ht]
		\includegraphics[width=0.9\columnwidth]{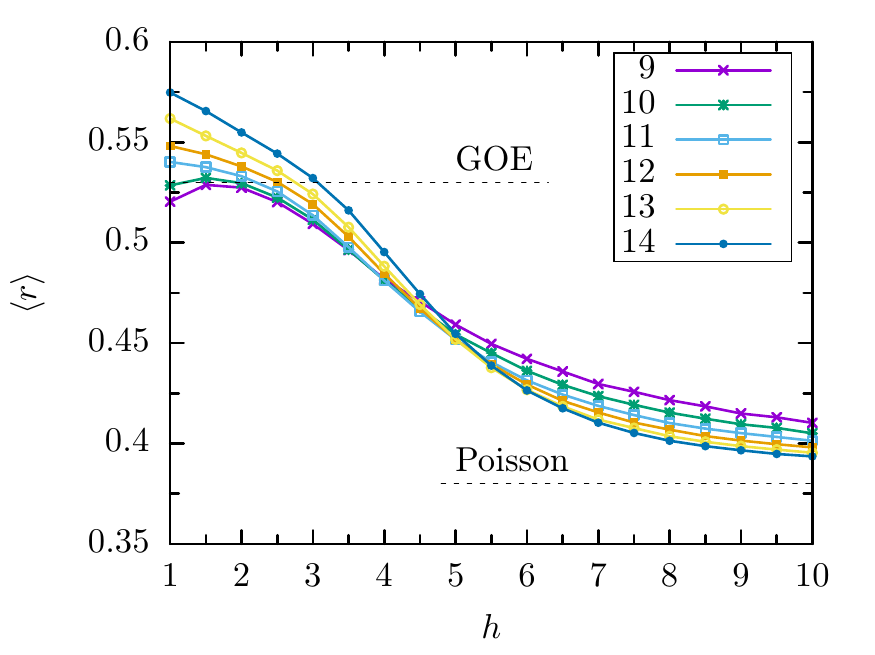}
		\caption{
	  	 \textcolor[rgb]{0.0,0.0,0.0}{
			Level statistics for the ferromagnetic nearest neighbor interaction $J_1 = -1$, as a function of the strength of disorder $h$ of the magnetic field $B^z \in \left\langle -h, h \right\rangle$. The DM interaction term is small $D=0.01$. The dominant symmetry of the system is GOE. The influence of GUE enhances with the system's size $L$.
		 }
		}
		\label{fig:DM0,01}
	\end{figure}
	\begin{figure}[ht]
		\includegraphics[width=0.9\columnwidth]{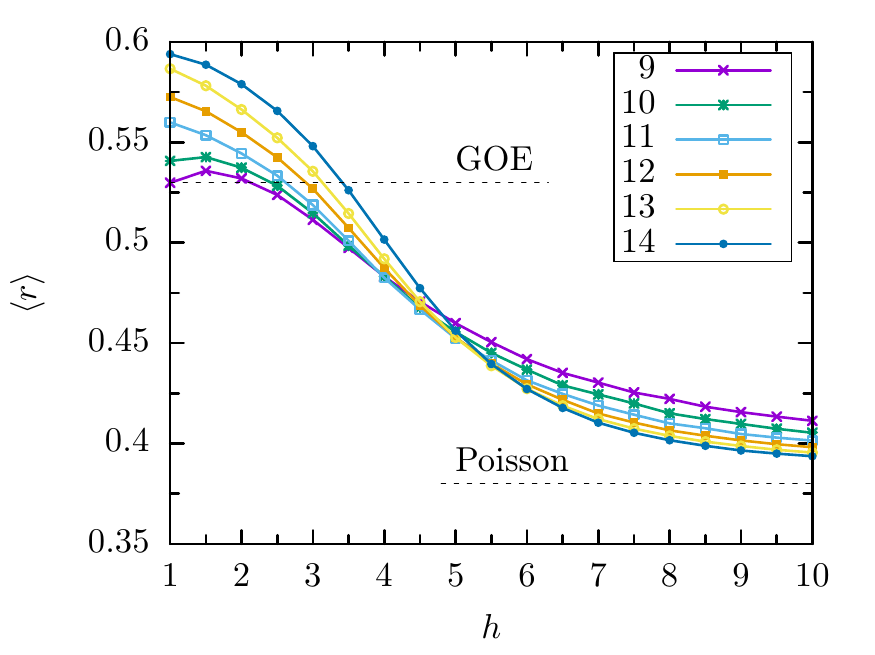}
		\caption{
		 \textcolor[rgb]{0.0,0.0,0.0}{
			The same as for Fig.~\ref{fig:DM0,01} but for $D=0.02$.
		 }
		}
		\label{fig:DM0,02}
	\end{figure}
	\begin{figure}[b]
		\includegraphics[width=0.9\columnwidth]{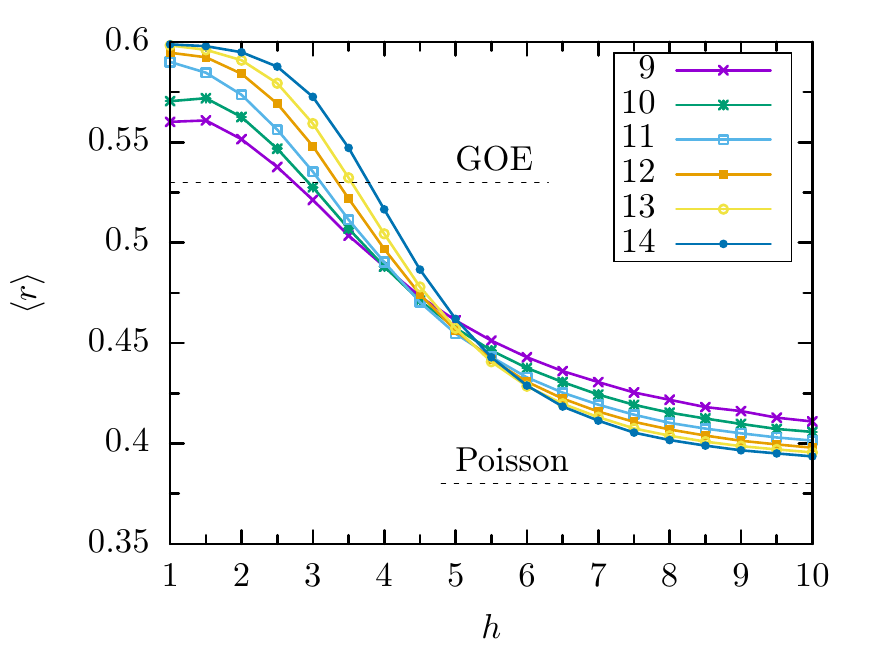}
		\caption{
		 \textcolor[rgb]{0.0,0.0,0.0}{
			Level statistics for  $J_1 = -1$, as a function of the strength of disorder $h$ of the magnetic field $B^z \in \left\langle -h, h \right\rangle$. The DM interaction term is stronger $D=0.05$. The symmetry of the system is mixed.
		 }
		}
		\label{fig:DM0,05}
	\end{figure}
	For a detailed analysis of the dependence on the critical strength of randomness $h_{c}$, we employ the scaling function \cite{Luitz} $g\left[L^{1/\nu}\left(h-h_{c}\right)\right]$ which allows collapsing all data to a single curve. Thus, for a given exponent $\nu$ we identify a critical disorder strength $h_{c}$. Fig.~\ref{GOE_GUE}b indicates a critical disorder strength of {$h_c = 4.7$} in the case of a ferromagnetic nearest neighbor interaction. The next-nearest neighbor antiferromagnetic spin interaction causes frustration and opens an extra channel for the energy redistribution in the system enhancing the critical disorder required for the MBL phase to {$h_c = 6.2$}. For ferromagnetic $J_{1}\neq 0$ and \textcolor[rgb]{0.0,0.0,0.0}{DM interaction} $D\neq 0$, and only $J_{2}=0$, the required disorder strength for MBL phase is noticeably decreased {$h_c = 4.4$}. It was proposed \cite{Kudo} that the transition to the MBL phase in the XXZ model occurs for $h>J_{1}$ with numerical results~\cite{GarciaGarcia} indicating $h>3.5J_{1}$. Implementing the scaling procedure, we infer  {$h_c = 4.6 J_{1}$}. As follows from Eq.~(\ref{Hamiltonian2}) for $J_{2}=0$ the XXZ model is retrieved with an effective XX constant $J^\prime_1 = \sqrt{J_1^{2}+D^{2}}/2$. For $D<J_{1}$ and $J^\prime_1 <J_{1}$, the XXZ system is gapped in line with the observation that  MBL  is reached at  weaker disorder. For $J_{2}\neq 0$ the equivalence with the XXZ model no longer holds and the DMI interaction extends the ergodic phase to  {$h_{c}=7.3$}.

	\textcolor[rgb]{0.0,0.0,0.0}{As Fig.~\ref{GOE_GUE} shows, the intrinsically complex nature of the Hamiltonian in the presence of a relatively strong \textcolor[rgb]{0.0,0.0,0.0}{DM interaction} causes a transition from a GOE to a GUE}. As clarified by the mapping to Eq.~\eqref{Hamiltonian2}, the complex part of the Hamiltonian is only due to the twisted boundary conditions. Hence, a gradual crossover from the GUE to the GOE with increasing $L$ is expected and \textcolor[rgb]{0.0,0.0,0.0}{confirmed by Fig.~\ref{GOE_GUE}c. For this reason, a finite-size scaling is applicable for a large enough disorder}.

\section{Quantum fluctuations and histograms of counts}\label{s:Qfhc}
\begin{figure}[!htb]
		\centering
		\includegraphics[width=0.99\columnwidth]{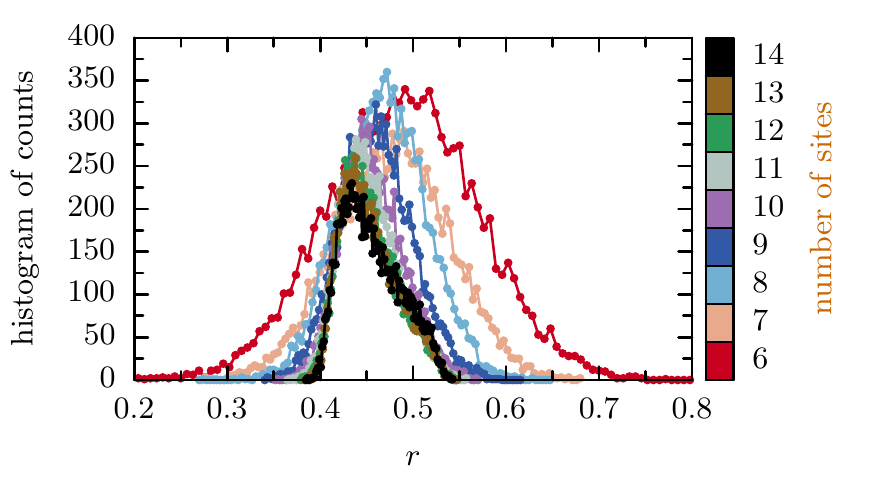}
		\caption{
	    	Histograms of counts for a fixed strength of disorder $h = 5$ without \textcolor[rgb]{0.0,0.0,0.0}{DM interaction} as a function of the consecutive level spacing $r$. Convergence is indicated for L=14.
		}
		\label{fig:brd_L}
	\end{figure}

\begin{figure}[!htb]
		\centering
		\includegraphics[width=0.98\columnwidth]{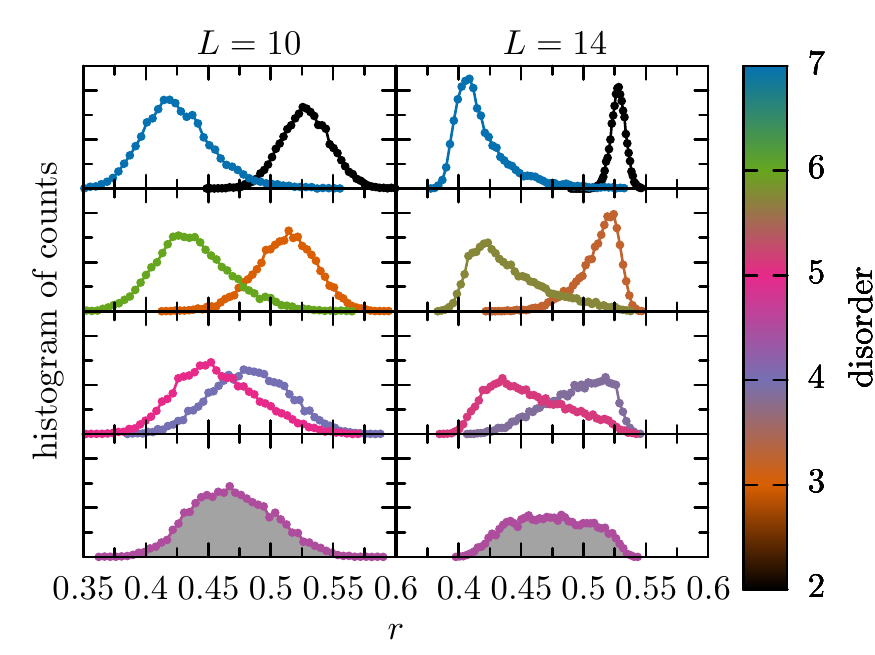}
		\caption{			
			Histograms of counts for the different strength of disorder $h$ with $J_1 = -1$ without \textcolor[rgb]{0.0,0.0,0.0}{DM interaction} as a function of the consecutive level spacing $r$. Broadening of the histogram corresponds to the critical strength of disorder and to the transition point. The larger is $L$ the peaks are more distinguished.
		}
		\label{fig:brd}
	\end{figure}

\begin{figure}[ht]
		\centering
		\includegraphics[width=0.7\columnwidth]{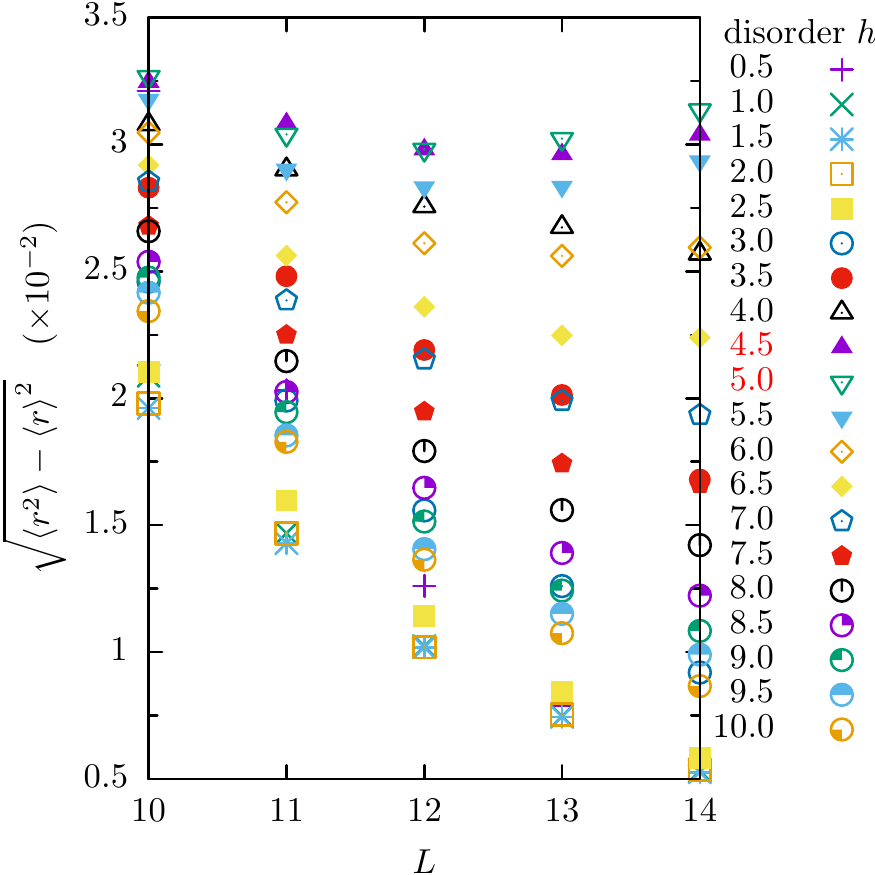}
		\caption{
			The fluctuation dependency on system size without $D$ and $J_1 = -1$. The two datasets marked by red are the nearest to critical disorder.
		}
		\label{fig:fluct}
	\end{figure}

	To formulate a possibly general criterion for MBL that is applicable in such cases, as well we analyzed the full statistics for each realization $\alpha$ of the random magnetic fields $r^{(\alpha)}$. The histograms corresponding to a counting classification of $r^{(\alpha)}$ for a given disorder strength $h$ is presented on Fig.~\ref{fig:brd}. As can be inferred, the histograms are narrow far away from the MBL transition, while the histograms become particularly broad close to the transition point mimicking the behavior of fluctuations near a conventional phase transitions. The histograms become more and more pronounced for increasing $L$.
	Fig.~\ref{fig:brd_L} demonstrates the convergence of the histograms of counts for the chains of different lengths. As we see already for $L=14$ counts histograms amalgamate underlying that an analysis of the histograms can serve as a further indicator in addition to  finite-size scaling. The convergence of  histograms even for relatively small systems endorses our method as less computationally demanding which is a major advantageous for exact diagonalization approaches that are considered as well suited for MBL studies.
	As for the  histograms of consecutive level spacing, Fig.~\ref{fig:brd} illustrates the broadening of the histograms when approaching the transition point between the ergodic and the MBL phases. As evident, the effect of broadening is even more prominent for systems with a larger size Fig.~\ref{fig:brd}. We note that the observed phenomena is not related to a particular type of level statistics but it is rather akin to the transition regime. Away from the transition point on the ergodic side (GOE statistics) and on the MBL phase side (Poisson statistics) the width of  histograms  are narrower. The broadening is linked to the enhanced quantum fluctuations Fig.~\ref{fig:fluct}.
	This behavior is of a general character and is maintained even after adding next nearest neighbor interaction and \textcolor[rgb]{0.0,0.0,0.0}{DM interaction} terms.

\begin{figure}[ht]
		\centering
		\includegraphics[width=0.98\columnwidth]{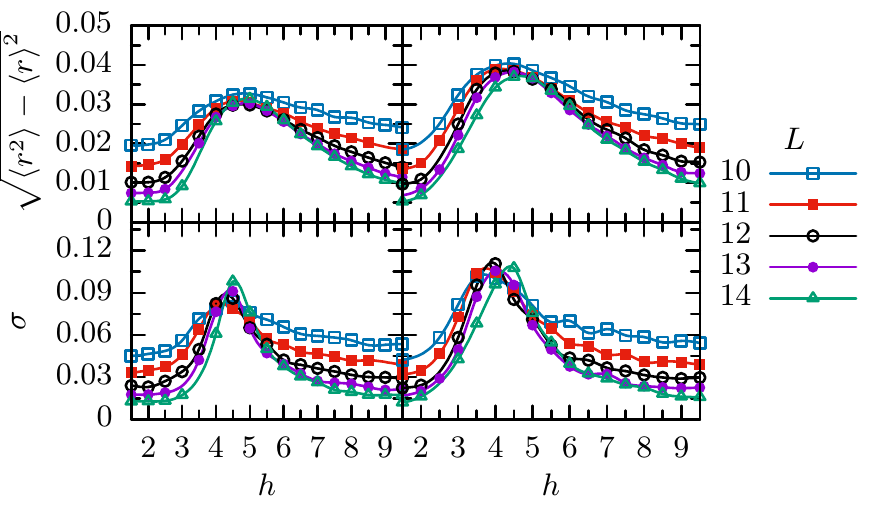}
		\caption{ Full width at half maximum $\sigma$ for the histograms  as a function of disorder $h$. The graphs on right side are with finite \textcolor[rgb]{0.0,0.0,0.0}{DM interaction}, $D = 0.2$.}
		\label{fig:fss}
	\end{figure}

	Physically, the broadening of  histograms  is attributable to the enhanced fluctuations near phase transitions (cf.~Figs.~\ref{fig:fluct}, \ref{fig:fss}). Hence, such broadening  serves as a further indicator for approaching the MBL phase.
	\begin{figure}[ht!]
		\centering
		\includegraphics[width=1.0\columnwidth]{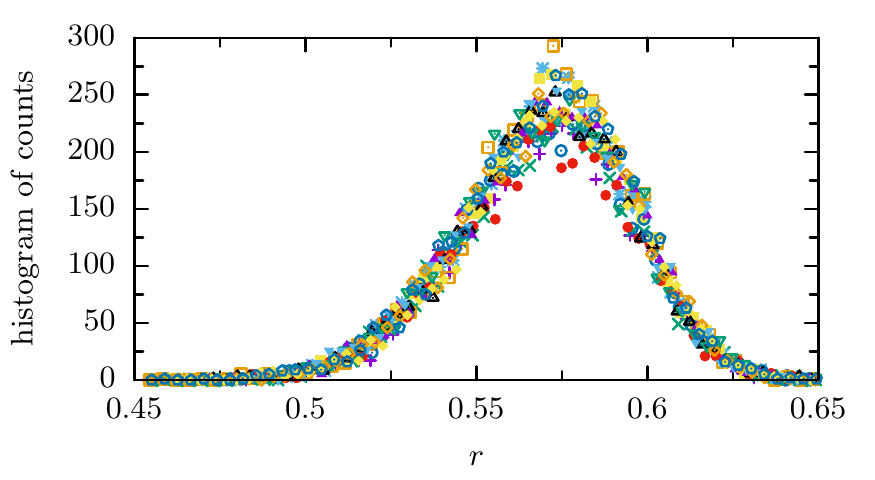}
		\caption{$J_1 = -1$, $D=0.2$, for fixed $h=5$. Additional site-dependent disorder of $10$\% in $D$ was taken while different colors and marks refer to various realizations.}
		\label{fig:d}
	\end{figure}

	Disorder in the exchange coupling or in $D$ may also  occur. The latter (cf.~Eq.~\ref{Hamiltonian}) can be  viewed as random change in $ \mathbf E$ or a random elastic energy change~(\textcolor[rgb]{0.0,0.0,0.0}{$ \mathbf E \cdot {\bf P} =g_{ME} \mathbf E \sum\limits^L_{i=1} \langle \mathbf{e}_x \times ( \mathbf{\hat S}_i \times \mathbf{\hat S}_{i+1} )\rangle $}),
	and thus, it is important for spin-phonon-coupled systems  at finite temperatures. Calculations evidence the robustness of the MBL phase against randomizing $D$ within a physically reasonable range, an example is depicted in  Fig.~\ref{fig:d}.

\textcolor[rgb]{0.0,0.0,0.0}{
	The results obtained for the different values of parameters are listed in the Table~\ref{tab:critical_h}.  Note, for  several particular values of the DM term which
	correspond to the mixed GOE/GUE statistics, the MBL phase can be identified through the present  method, while a finite size scaling procedure fails.
}
	\begin{table}[h]
		\centering
		\begin{tabular}{|rc|ccc|}
		\hline
		  $J_2$          & $D$            & $h_c$ & $h_\sigma$ & $h_{f}$ \\\hline
		                 &                & $4.6$ & $4.5$      & $4.8$   \\
		  $\frac{1}{4}$  &                & $6.2$ & $5.3$      & $5.7$   \\
		  $-\frac{1}{4}$ &                & $5.4$ & $4.6$      & $5.2$   \\
		                 & $\frac{1}{5}$  & $4.4$ & $4.4$      & $4.7$   \\
		  $\frac{1}{4}$  & $\frac{1}{5}$  & $7.3$ & $5.0$      & $5.5$   \\
		  $-\frac{1}{4}$ & $\frac{1}{5}$  & $6.4$ & $4.9$      & $5.1$   \\\hline
		\end{tabular}
			\begin{tabular}{|rc|ccc|}
			\hline
		 $J_2$           & $D$            & $h_c$ & $h_\sigma$ & $h_{f}$ \\\hline
			             &                & $4.7$ & $4.5$      & $4.8$   \\
			             & $\frac{1}{100}$& $5.0$ & $4.5$      & $4.8$   \\
		 	             & $\frac{1}{50}$ & $5.3$ & $4.4$      & $4.6$   \\
		 	             & $\frac{1}{20}$ & $5.6$ & $4.0$      & $4.5$   \\ 	
		 	             & $\frac{1}{10}$ & $5.4$ & $4.2$      & $4.6$   \\ 	
		 	             & $\frac{1}{5}$  & $4.3$ & $4.4$      & $4.7$   \\ 	  	
			\hline
			\end{tabular}
		\caption{
		 \textcolor[rgb]{0.0,0.0,0.0}{
		  The estimated critical disorders $h$ with $J_1 = -1$, a) $h_c$ from a scaling procedure for $\left\langle r \right\rangle$, b) $h_\sigma$ from an analysis of the full-width at half-height of the histograms, c) $h_{f}$ from the analysis of the the fluctuations $\sqrt{\left\langle r^2 \right\rangle - \left\langle r \right\rangle^2 }$. The left dataset with $L=\{ 10, 12, 14 \}$ and $4096$ realizations for each mean of the consecutive level spacings $\left\langle r \right\rangle$. The right dataset was prepared with $L=\{ 10, 11, 12, 13, 14 \}$ and $10240$ realizations.
		 }
		}
		\label{tab:critical_h}
	\end{table}
	
\section{Conclusions}
	Summarizing, our numerics and analysis evidence  MBL phase in a chiral multiferroic chain. A new,  general indicator for approaching the MBL phase  is identified and tested  against  conventional procedures. In view of the experimental feasibility of  materials and settings, the predictions might have some  signatures experimentally,   for instance through investigating the transport and excitation spectrum of electromagnons.
	
\section{Acknowledgment}
	We thank Marcus Heyl for fruitful discussions and numerous comments on the manuscript. 
	We acknowledge financial support from DFG through SFB 762, and BE 2161/5-1.

\appendix

\section{Scaling procedure}
	In Ref.~[\onlinecite{Luitz}] a systematic analysis of the fitting windows and system sizes in relation to the critical disorder $h_c$ and $\nu$ was performed. For the histograms  windowing of the energy spectrum from $S^z = 0$ block causes equalizing disproportions of the histograms from different phases. Therefore, the full spectrum of the biggest block is considered. Although, for a bigger system size the middle part of the spectrum can be successfully used. The level statistics of $\left\langle r \right\rangle$ as a function of disorder $h$ are scaled with $f_L(h) = L^{1/\nu} (h-h_c)$, where $\nu = 1$ was assumed.
	For best fitting parameter $h_c$, the scaling procedure can supported by a minimizing function $w(h)$
	\begin{equation}
		 w = \sum\limits_{L, L^\prime} \int\limits_{h_1}^{h_2} \left| \left\langle r \right\rangle(f_L(h)) - \left\langle r\right\rangle(f_{L^\prime}(h))  \right| d h
		 \label{eq:min}
	\end{equation}
	where $h_1$ and $h_2$ are defined by the common integration domain
	$	h_1 = \max \left( L_i^{\frac{1}{\nu}} (a - h_c) \right),\;
		h_2 = \min \left( L_i^{\frac{1}{\nu}} (b - h_c) \right)$
	were $L_i$ denote the system sizes to be analyzed and $a = 1$, $b = 10$ are boundaries limit for unscaled data set.

	The critical disorder also can be inferred by minimizing the distances of the peaks positions for $\left\langle r \right\rangle$ first derivatives, which is more accurate than minimizing global overlaps, especially with non-zero DM interaction. With Eq.~(\ref{eq:min}) a proper adjusting of integration limits is required.


\begin{thebibliography}{99}
	\bibitem{Anderson}
		P.~W.~Anderson,
		Phys. Rev. \textbf{109}, 1492 (1958),
		\doiit{10.1103/PhysRev.109.1492}
		
	\bibitem{wave1}
		M.~Segev, Y.~Silberberg, D.~N.~Christodoulides,
		Nat. Photonics \textbf{7}, 197 (2013),
		\doiit{10.1038/nphoton.2013.30}
		
	\bibitem{wave2}
		H.~Hu, A.~Strybulevych,  J.~Page, S.~Skipetrov, and B.~Van~Tiggelen,
		Nature Phys. \textbf{4}, 945–948 (2008),
		\doiit{10.1038/nphys1101}
		
	\bibitem{wave3}
		A.~Lagendijk, B.~van~Tiggelen, and D.~S.~Wiersma,
		Phys. Today \textbf{62}, 24–29 (2009),
		\doiit{10.1063/1.3206091}
		
\bibitem{eeanderson}
	\textit{Electron-electron Interactions in Disordered Systems}, Eds. M.~Pollak and A.~L.~Efros (North-Holland, Amsterdam, 1984).
		
	\bibitem{Basko}
		D.~M.~Basko, I.~L.~Aleiner, B.~L.~Altshuler,
		Annals of Physics \textbf{321}, 1126 (2006),
		\doiit{10.1016/j.aop.2005.11.014}
	
	\bibitem{Devakul}
		L.~Zhang, B.~Zhao, T.~Devakul, D.~A.~Huse,
		Phys. Rev. B \textbf{93}, 224201 (2016),
		\doiit{10.1103/PhysRevB.93.224201}
		
	\bibitem{Luitz}
		D.~J.~Luitz, N.~Laflorencie, and F. Alet,
		Phys. Rev. B \textbf{91}, 081103(R) (2015),
		\doiit{10.1103/PhysRevB.91.081103}
		D.~J.~Luitz, N.~Laflorencie, and F. Alet,
		Phys. Rev. B \textbf{93}, 060201(R) (2016),
		\doiit{10.1103/PhysRevB.93.060201}
		
	\bibitem{Vasseur}
		R.~Vasseur, A.~J.~Friedman, S.~A.~Parameswaran, and A.~C.~Potter,
		Phys. Rev. B \textbf{93}, 134207 (2016).
		\doiit{10.1103/PhysRevB.93.134207}
		
	\bibitem{Bardarson}
		J.~A.~Kj\"all, J.~H.~Bardarson, and F.~Pollmann,
		Phys. Rev. Lett. \textbf{113}, 107204 (2014),
		\doiit{10.1103/PhysRevLett.113.107204}
	
	\bibitem{Serbyn}
		M.~Serbyn and J.~E.~Moore,
		Phys. Rev. B \textbf{93}, 041424(R) (2016),
		\doiit{10.1103/PhysRevB.93.041424}
		
	\bibitem{topical} For an overview we refer to the topical issue {\it Many-Body Localization}  ed. J. H. Bardarson {\it  et al.} Ann.~Phys.  (2017).
	
	\bibitem{huse}
		V.~Oganesyan, A.~Pal, D.~A.~Huse,
		Phys. Rev. B \textbf{80}, 115104 (2009),
		\doiit{10.1103/PhysRevB.80.115104}
		
	\bibitem{mblatoms1}
		M.~Schreiber {\it et al.},
		Science \textbf{349}, 842-845 (2015),
		\doiit{10.1126/science.aaa7432}
		
	\bibitem{mblatoms2}
		Jae-yoon~Choi {\it et al.},
		Science \textbf{352}, 1547-1552 (2016),
		\doiit{10.1126/science.aaf8834}
		
	\bibitem{mblatoms3}
		S.~S.~Kondov, {\it et al.},
		Phys. Rev. Lett. \textbf{114}, 083002 (2015),
		\doiit{10.1103/PhysRevLett.114.083002}
		
	\bibitem{mblions}
		J.~Smith {\it et al.},
		Nature Physics \textbf{12}, 907–911 (2016),
		\doiit{10.1038/nphys3783}
		
	\bibitem{chiral}
		H.~Katsura, N.~Nagaosa, and A.~V.~Balatsky,
		Phys. Rev. Lett. \textbf{95}, 057205 (2005),
		\doiit{10.1103/PhysRevLett.95.057205}
		M.~Mostovoy,
		Phys. Rev. Lett. \textbf{96}, 067601 (2006),
		\doiit{10.1103/PhysRevLett.96.067601}
		I.~A.~Sergienko and E.~Dagotto,
		Phys. Rev. B \textbf{73}, 094434 (2006),
		\doiit{10.1103/PhysRevB.73.094434}
		Y.~Tokura, S.~Seki, and N.~Nagaosa,
		Rep. Prog. Phys. \textbf{77}, 076501 (2014),
		\doiit{10.1088/0034-4885/77/7/076501}
		Y.~Tokura and S.~Seki,
		Adv. Mater. \textbf{22}, 1554 (2010),
		\doiit{10.1002/adma.200901961}
		K.~F.~Wang, J.~M.~Liu, and Z.~F. Ren,
		Adv. Phys. \textbf{58}, 321 (2009),
		\doiit{10.1080/00018730902920554}
	
	\bibitem{bo}
		H.-B.~Chen, Y.-Q.~Li, and J.~Berakdar,
		J. Appl. Phys. \textbf{117}, 043910 (2015),
		\doiit{10.1063/1.4906520}
		
	\bibitem{Bertini}
		B.~Bertini, F.~H.~L.~Essler, S.~Groha, and N.~J.~Robinson,
		Phys. Rev. Lett. \textbf{115}, 180601 (2015),
		\doiit{10.1103/PhysRevLett.115.180601}
		
	\bibitem{Rigol}
		M.~Rigol, V.~Dunjko, V.~Yurovsky, and M.~Olshanii,
		Phys. Rev. Lett. \textbf{98}, 050405 (2007),
		\doiit{10.1103/PhysRevLett.98.050405}
	
	\bibitem{Ilievski}
		E.~Ilievski, J.~De~Nardis, B.~Wouters, J.-S.~Caux, F.~H.~L.~Essler, and T.~Prosen,
		Phys. Rev. Lett. \textbf{115}, 157201 (2015),
		\doiit{10.1103/PhysRevLett.115.157201}
		G.~Misguich, V.~Pasquier, and J.~M.~Luck,
		Phys. Rev. B \textbf{94}, 155110 (2016),
		\doiit{10.1103/PhysRevB.94.155110}

	\bibitem{Pozsgay}
		B.~Pozsgay, M.~Mestyán, M.~A.~Werner, M.~Kormos, G.~Zaránd, and G.~Takács,
		Phys. Rev. Lett. \textbf{113}, 117203 (2014),
		\doiit{10.1103/PhysRevLett.113.117203}

	\bibitem{Mierzejewski}
		M.~Mierzejewski, P.~Prelovšek, and T.~Prosen,
		Phys. Rev. Lett. \textbf{113}, 020602 (2014),
		\doiit{10.1103/PhysRevLett.113.020602}
%
	\bibitem{ETH1}
		J.~Eisert, M.~Friesdorf and C.~Gogolin,
		Nat. Phys. \textbf{11}, 124 (2015),
		\doiit{10.1038/nphys3215}
		
	\bibitem{ETH2}
		C.~Gogolin and J.~Eisert,
		Rep. Prog. Phys. \textbf{79},  056001 (2016),
		\doiit{10.1088/0034-4885/79/5/056001}
		
	\bibitem{ETH3}
		L.~D’Alessio, Y.~Kafri, A.~Polkovnikov, and M.~Rigol,
		Adv. Phys. \textbf{65}, 239–362 (2016),
		\doiit{1080/00018732.2016.1198134}
		
	\bibitem{huse1}
		M.~Serbyn, Z.~Papic, and D.~A.~Abanin,
		Phys. Rev. Lett. \textbf{111}, 127201 (2013),
		\doiit{10.1103/PhysRevLett.111.127201}
		
	\bibitem{huse2}
		D.~A.~Huse and V.~Oganesyan,
		Phys. Rev. B \textbf{90}, 174202 (2014),
		\doiit{10.1103/PhysRevB.90.174202}
		
	\bibitem{Bauer}
		B.~Bauer and C.~Nayak,
		J. Stat. Mech.: Theor. Exp. \textbf{P09005} (2013),
		\doiit{10.1088/1742-5468/2013/09/P09005}
		V.~Oganesyan and D.~A.~Huse,
		Phys. Rev. B \textbf{75}, 155111 (2007),
		\doiit{10.1103/PhysRevB.75.155111}
		A.~Pal and D.~A.~Huse,
		Phys. Rev. B \textbf{82}, 174411 (2010),
		\doiit{10.1103/PhysRevB.82.174411}
		C.~R.~Laumann, A.~Pal, and A.~Scardicchio,
		Phys. Rev. Lett. \textbf{113}, 200405 (2014),
		\doiit{10.1103/PhysRevLett.113.200405}
		
	\bibitem{Haake}
		F.~Haake, Quantum Signatures of Chaos (Springer, Berlin, 2000).
	
	\bibitem{Bera}
		S. Schierenberg, F. Bruckmann, and T. Wettig Phys. Rev. E \textbf{85}, 061130 (2012),
		
		A.~Rycerz,
		Phys. Rev. B \textbf{85}, 245424 (2012),
		\doiit{10.1103/PhysRevB.85.245424}
		S.~Schierenberg, F.~Bruckmann, and T.~Wettig,
		Phys. Rev. E \textbf{85}, 061130 (2012),
		\doiit{10.1103/PhysRevE.85.061130}
		
	\bibitem{Park}
		S.~Park, Y.~J.~Choi, C.~L.~Zhang, and S-W.~Cheong,
		Phys. Rev. Lett. \textbf{98}, 057601 (2007),
		\doiit{10.1103/PhysRevLett.98.057601}
		
	\bibitem{Schrettle}
		F.~Schrettle, S.~Krohns, P.~Lunkenheimer, J.~Hemberger, N.~Büttgen, H.-A.~Krug von Nidda, A.~V.~Prokofiev, and A.~Loidl,
		Phys. Rev. B \textbf{77}, 144101 (2008),
		\doiit{10.1103/PhysRevB.77.144101}
		
	\bibitem{Azimi}
		M.~Azimi, L.~Chotorlishvili, S.~K.~Mishra, S.~Greschner, T.~Vekua, and J.~Berakdar,
		Phys. Rev. B \textbf{89}, 024424 (2014),
		\doiit{10.1103/PhysRevB.89.024424}
		
	\bibitem{Majumdar}
		C.~K.~Majumdar and D.~K.~Ghosh,
		J. Math. Phys. \textbf{10}, 1388 (1969),
		\doiit{10.1063/1.1664978}
		A.~V.~Chubukov,
		Phys. Rev. B \textbf{44}, 4693 (1991),
		\doiit{10.1103/PhysRevB.44.4693}
		R.~Bursill, G.~A.~Gehring, D.~J.~J.~Farnell, J.~B.~Parkinson, T.~Xiang, and C.~Zeng,
		J. Phys.: Condens. Matter \textbf{7}, 8605 (1995),
		\doiit{10.1088/0953-8984/7/45/016}
		
	\bibitem{Kecke2008}
		T.~Hikihara, L.~Kecke, T.~Momoi, and A.~Furusaki,
		Phys. Rev. B {\bf 78}, 144404 (2008),
		\doiit{10.1103/PhysRevB.78.144404}
		
	\bibitem{Sirker}
		J.~Sirker, Z.~Weihong, O.~P.~Sushkov, and J.~Oitmaa,
		Phys. Rev. B \textbf{73}, 184420 (2006),
		\doiit{10.1103/PhysRevB.73.184420}
		
	\bibitem{jia1}
		C.~Jia and J.~Berakdar,
		Eur. Phys. Lett. \textbf{85}, 57004 (2009),
		\doiit{10.1209/0295-5075/85/57004}
		
	\bibitem{jia2}
		C.~Jia, J.~Berakdar,
		Phys. Stat. Sol. B \textbf{247}, 662 (2010),
		\doiit{10.1002/pssb.200983028}
		
	\bibitem{a}
		E.~Levi, M.~Heyl, I.~Lesanovsky, and J.~P.~Garrahan,
		Phys. Rev. Lett. \textbf{116}, 237203 (2016),
		\doiit{10.1103/PhysRevLett.116.237203}
		
	\bibitem{b}
		M.~H.~Fischer, M.~Maksymenko, and E.~Altman,
		Phys. Rev. Lett. \textbf{116}, 160401 (2016),
		\doiit{10.1103/PhysRevLett.116.160401}
		
	\bibitem{c}
		M.~V.~Medvedyeva, T.~Prosen, and M.~Znidaric,
		Phys. Rev. B \textbf{93}, 094205 (2016),
		\doiit{10.1103/PhysRevB.93.094205}
		
	\bibitem{d}
		H.~P.~Lüschen {\it et al. },
		Phys. Rev. X \textbf{7}, 011034 (2017),
		\doiit{10.1103/PhysRevX.7.011034}
		
	\bibitem{stm}
		M.~Menzel, {\it et al.},
		Phys. Rev. Lett. \textbf{108}, 197204 (2012),
		\doiit{10.1103/PhysRevLett.108.197204}
		Phys. Rev. Lett. \textbf{112}, 047204 (2014),
		\doiit{10.1103/PhysRevLett.112.047204}
		D.~Serrate, {\it et al.} Nature Nanotechnology \textbf{5}, 350 (2010),
		\doiit{10.1038/nnano.2010.64}
		
	\bibitem{Bocquet}
		M.~Bocquet, F.~H.~L.~Essler, A.~M.~Tsvelik, and A.~O.~Gogolin,
		Phys. Rev. B \textbf{64}, 094425 (2001),
		\doiit{10.1103/PhysRevB.64.094425}
		
	\bibitem{Kudo}
		K.~Kudo, T.~Deguchi,
		Phys. Rev. B \textbf{69}, 132404 (2004),
		\doiit{10.1103/PhysRevB.69.132404}
		
	\bibitem{GarciaGarcia}
		C.~L.~Bertrand, A.~M.~Garcia-Garcia,
		Phys. Rev. B \textbf{94}, 144201 (2016),
		\doiit{10.1103/PhysRevB.94.144201}
\end{thebibliography}
\end{document}